\documentclass{article}
\usepackage{spconf,amsmath,graphicx}
\usepackage{graphicx}
\usepackage{subfigure}
\usepackage{multirow}
\usepackage{array}
\usepackage{footnote}

\usepackage{algorithmic,algorithm}

\usepackage{dsfont}
\usepackage{amsfonts,amsmath}
\usepackage{url,cite}
\usepackage{color}

\usepackage{threeparttable,balance}

\def\cred{\textcolor{black}}  

\def\X{{\mathbf X}}

\title{Hyperspectral image
super-resolution with deep priors \\ and degradation model inversion}
	
\name{Author(s) Name(s)\thanks{Thanks to XYZ agency for funding.}}
\name{Xiuheng Wang$^*$, Jie Chen$^\dagger$, C\'edric Richard$^*$}
\address{\small  $^*$ Universit\'e  C\^ote d'Azur, CNRS, OCA, France\\
	\small $^\dagger$ School of Marine Science and Technology, Northwestern Polytechnical University, China \\
	\small xiuheng.wang@oca.eu, dr.jie.chen@ieee.org, cedric.richard@unice.fr}
\begin{document}
\ninept
\maketitle
\begin{abstract}
To overcome inherent hardware limitations of hyperspectral imaging systems with respect to their spatial resolution, fusion-based hyperspectral image (HSI) super-resolution is attracting increasing attention. This technique aims to fuse a low-resolution (LR) HSI and a conventional high-resolution (HR) RGB image in order to obtain an HR HSI. Recently, deep learning architectures have been used to address the HSI super-resolution problem and have achieved remarkable performance. However, they ignore the degradation model even though this model has a clear physical interpretation and may contribute to improve the performance. We address this problem by proposing a method that, on the one hand, makes use of the linear degradation model in the data-fidelity term of the objective function and, on the other hand, utilizes the output of a convolutional neural network for designing a deep prior regularizer in spectral and spatial gradient domains. Experiments show the performance improvement achieved with this strategy.
\end{abstract}

\section{Introduction}
\label{sec:introduction}

Hyperspectral imaging systems collect images of a scene using contiguous spectral bands ranging from ultraviolet to visible and infrared. Hyperspectral imaging is beneficial in applications as diverse as remote surveillance, medicine, and environmental monitoring. Unfortunately, many factors such as noise and low sensing resolution can cause image degradation during the acquisition process. Image restoration methods are often employed before further processing. Recently, it has become popular to restore an HR HSI by fusing an LR HSI and an HR RGB image. This framework, often called \emph{fusion-based HSI super-resolution}~\cite{lanaras2015hyperspectral}, raises the challenge of accounting correlation between spectral bands while ensuring spatial consistency~\cite{henrot2012fast}.

In the scope of this work, we consider the problem of constructing an HR HSI $\mathcal{X}\in\mathbb{R}^{B\times L \times W}$ via the fusion of an LR HSI $\mathcal{Y}\in\mathbb{R}^{B\times l \times w}$ and an HR RGB image $\mathcal{Z}\in\mathbb{R}^{b\times L \times W}$, where $B$ and $b$ are the numbers of spectral channels of the hyperspectral image and the RGB image, respectively, with $B > b$, and $(L \times W)$ and $(l \times w)$ denote the dimensions of the HR image and the LR image, respectively, with $L > l$ and $W > w$. Images $\mathcal{X}$, $\mathcal{Y}$ and $\mathcal{Z}$ can be reshaped in matrix forms $\mathbf{X}\in\mathbb{R}^{B\times N}, \mathbf{Y}\in\mathbb{R}^{B\times n}$ and $\mathbf{Z}\in\mathbb{R}^{b\times N}$, respectively, where $N = L \times W$ and $n = l \times w$ are the numbers of pixels in each band of the HR and LR images. Considering a linear degradation model, $\mathbf{Y}$ and $\mathbf{Z}$ can be interpreted as down-sampled versions of $\mathbf{X}\in\mathbb{R}^{B\times N}$ in the spatial and spectral domains, respectively:
\begin{equation}
    \label{eq:degradation}
    \mathbf{Y} = \mathbf{XBS},\quad
    \mathbf{Z} = \mathbf{RX},
\end{equation}
where $\mathbf{B}\in\mathbb{R}^{N\times N}$ is the spatial blurring matrix, $\mathbf{S}\in\mathbb{R}^{N\times n}$ is a downsampling operator with factor $s=L/l$, and $\mathbf{R}\in\mathbb{R}^{b\times B}$ denotes the spectral response function (SRF) of the RGB camera.
The HSI super-resolution problem can be formulated as the estimation of ${\bf X}$ given the observed data ${\bf Y}$ and ${\bf Z}$, with known system responses $\mathbf{B}$, $\mathbf{S}$ and~$\mathbf{R}$. Based on~\eqref{eq:degradation}, this problem can be written as the minimization of an unconstrained objective function of the form:
\begin{equation}
    \label{eq:object_0}
    J(\mathbf{X}) =
    \|\mathbf{Y} - \mathbf{XBS}\|_{F}^{2} +
    \|\mathbf{Z} - \mathbf{RX}\|_{F}^{2} +
    \varphi(\mathbf{X})
\end{equation}
where $\varphi(\cdot)$ is some regularization functions. 
Reconstructing $\bf X$ from $\bf Y$ and $\bf Z$ by minimizing~\eqref{eq:object_0} without $\varphi(\cdot)$ is a highly ill-posed problem. This justifies the use of $\varphi(\cdot)$ to constrain the solution space by promoting prior information on $\bf X$. 

Under the classic optimization framework, handcrafted forms of  $\varphi(\cdot)$, which promote the sparsity, spatial continuity and edge preserving, have been intensively studied for this problem~\cite{lanaras2015hyperspectral, wei2016multiband, akhtar2014sparse, dong2016hyperspectral, dian2019learning}, and also for other highly related problems~\cite{chen2013nonlinear, song2019online}.
However, designing a powerful $\varphi(\cdot)$ is not trivial and may also cause difficulty in finding optimal solutions.
Inspired by the success of deep learning in computer vision, convolutional neural networks (CNN) have been used to get $\bf X$ from the fusion of $\bf Y$ and $\bf Z$~\cite{palsson2017multispectral, yang2017pannet,xie2019multispectral, zhang2020unsupervised}. Deep learning methods require less handcrafted prior information
on $\bf X$ and have been shown to achieve significant performance enhancement compared to model-based methods such as~\eqref{eq:object_0}. However, they need massive data for training and may not be consistent with the physical degradation model of form~\eqref{eq:degradation} involving $\bf Y$ and $\bf Z$. A brief review of these methods is given in Section~\ref{sec:work}.

\begin{figure}[tp]
	\centering
	\includegraphics[scale=0.3]{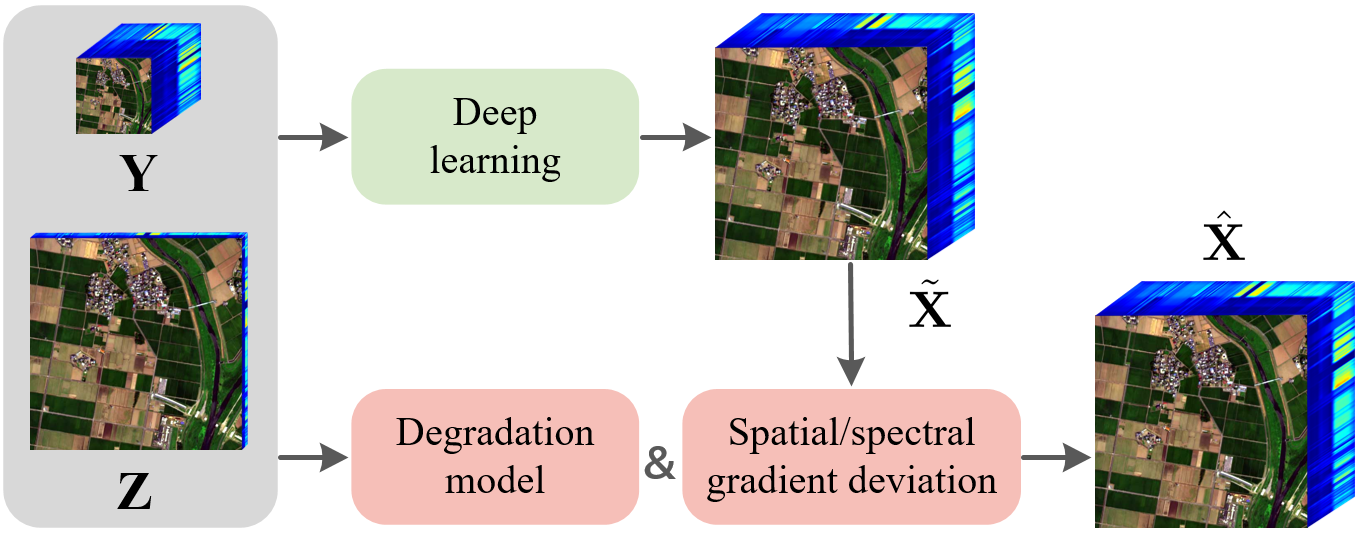}
	\caption{HSI super-resolution with deep priors and model degradation inversion accounting for spatial/spectral gradient deviation of HSIs.
	\vspace{-3mm}}
	\label{fig:scheme}
\end{figure}

To leverage the merits of both model-based and deep learning methods, recent approaches have started to plug the output of a CNN, denoted as $\tilde{\bf X}$, into the objective function~\eqref{eq:object_0} as a \emph{deep prior regularizer}~\cite{xie2019hyperspectral, wang2021hyperspectral, vella2021enhanced}. Specially, the Frobenius norm $\varphi(\mathbf{X})=\|\mathbf{X} - \tilde{\bf X}\|_{F}^{2}$ is considered in~\cite{xie2019hyperspectral, wang2021hyperspectral}. It allows the use of a fast off-the-shelf solver. The 2D Total Variation (TV) norm $\varphi(\mathbf{X})=\|\mathbf{X} - \tilde{\bf X}\|_{TV}^{2}+\|\mathbf{X}\|_{TV}$ is used in~\cite{vella2021enhanced}. It is
applied to each band independently from the others to enforce smoothness in the spatial domain. Nevertheless, none of these methods simultaneously exploits the spectral-spatial gradient information for enhancing fusion process. 

The aim of this paper is to introduce a novel strategy for HSI super-resolution that, on the one hand, makes use of the physical linear degradation model in the data-fidelity term of the objective function and, on the other hand, exploits the spectral-spatial gradient difference of HSIs using a deep prior regularizer from the output of an CNN. Experimental results show the performance improvement achieved with this strategy.

\textit{Notation:} $\mathcal{X}$, $\mathbf{X}$ and $\mathbf{x}$ refer to the same 3D image~$\mathcal{X}$:  
matrix $\mathbf{X}$ is obtained by arranging the pixel column vectors in $\mathcal{X}$ next to each other; vector $\mathbf{x}$ is obtained by stacking the columns of $\mathbf{X}$ on top of each other. This notation system also works for other images.

\section{Related works}
\label{sec:work}
In this section, we provide a brief overview on some existing methods to facilitate the presentation of our method.
 
\vspace{-3mm}
\subsection{Model-based methods}
\vspace{-1mm}
Solving HSI super-resolution and unmixing in a joint framework has been demonstrated to significantly improve the super-resolution performance. In~\cite{lanaras2015hyperspectral}, the authors perform a joint unmixing of ${\bf Y}$ and ${\bf Z}$ that allows them to split the initial optimization problem into two constrained quadratic sub-problems, which can be solved efficiently. In~\cite{wei2016multiband}, the authors reconstruct the latent ${\bf X}$ by using a variable splitting technique \cred{w.r.t.} its endmembers and their corresponding abundances. Considering the similarity between neighboring pixels, the authors in~\cite{akhtar2014sparse} enforce group-sparsity and non-negativity properties in small image cubes. In~\cite{dong2016hyperspectral}, the authors introduce a non-negative sparse coding method to exploit the sparsity of pixels and the non-local spatial similarity of the latent ${\bf X}$. Another strategy in HSI super-resolution consists of tensor-based factorization. For instance, a graph Laplacian-guided coupled tensor decomposition model is proposed by the authors in~\cite{dian2019learning} to exploit the spatial-spectral information of HSI and RGB images.

\vspace{-3mm}
\subsection{Deep learning methods}
\vspace{-1mm}
Recently, deep learning has been proved to be an effective data-driven technique for solving HSI super-resolution problem. In~\cite{palsson2017multispectral}, the authors design an CNN with 3D convolution to fuse ${\bf Y}$ and ${\bf Z}$ and obtain ${\bf X}$ in an end-to-end manner. The PanNet architecture is proposed in~\cite{yang2017pannet} to preserve spectral and spatial information in HSI super-resolution problem. In~\cite{xie2019multispectral}, the authors unfold an iterative algorithm into a deep network called MHF-net. This algorithm is based on a novel LR/HR fusion model which takes the degradation models of ${\bf Y}$ and ${\bf Z}$ as well as the low-rankness of ${\bf X}$ into consideration. In~\cite{zhang2020unsupervised}, a two-stage network based on unsupervised adaption learning is proposed to learn priors of ${\bf X}$ while estimating the unknown spatial degradation.

\section{The proposed Method}
\label{sec:problem}

Considering the HSI super-resolution problem defined by the unconstrained objective function~\eqref{eq:object_0}, we propose to jointly consider the super-resolution from the physical model, i.e., the first two terms in (2), and the data-driven prior information denoted by $\tilde{\mathbf{X}}$. In this work, a prior image from the output of an CNN is used to enhance the physical model result via $\varphi(\mathbf{X})$ by reducing the difference between the optimization variable $\mathbf{X}$ and $\tilde{\mathbf{X}}$ in spectral and spatial gradient domains respectively. Under this design, the objective function~\eqref{eq:object_0} becomes:
\begin{equation}
    \label{eq:object}
    \begin{split}
    &J({\bf{X}})=\|\mathbf{Y} - \mathbf{XBS}\|_{F}^{2} +
 \|\mathbf{Z} - \mathbf{RX}\|_{F}^{2} + \varphi(\mathbf{X})\\
  &\rm{with}\,\, \varphi(\mathbf{X}) =  \mu\|\mathbf{D}(\mathbf{x} - \tilde{\mathbf{x}})\|^{2} + \nu\|\mathbf{E}(\mathbf{x} - \tilde{\mathbf{x}})\|^{2} \\
 &\rm{and}\qquad\ \, \tilde{\mathbf{X}} = \texttt{CNN}(\mathbf{Y}, \mathbf{Z})
    \end{split}
\end{equation}
where $\mathbf{x}\in\mathbb{R}^{BN}$ and $\tilde{\mathbf{x}}\in\mathbb{R}^{BN}$ denote the vectors obtained by stacking the columns of the matrices $\bf{X}$ and $\tilde{\bf{X}}$, respectively and $\texttt{CNN}$ denotes a trained CNN with the inputs $\mathbf{Y}$ and $\mathbf{Z}$ to produce $\tilde{\mathbf{X}}$. 
The first two terms of $J(\X)$ guarantee that the candidate solution is consistent with the degradation model~\eqref{eq:degradation}. The last term of $J(\X)$ are regularization terms used to promote prescribed properties, with positive hyper-parameters $\mu$ and $\nu$. These two properties consist of smooth error maps along the spatial and spectral dimensions, obtained with matrices $\mathbf{D}$ and $\mathbf{E}$ defined as follows.

Matrix $\mathbf{D}$ can be designed by choosing first a convolution kernel~\cred{$\mathcal{D}$.} We consider the Laplacian filter \cred{$\mathcal{D}_{\ell}$ for each channel $\ell$}:
\begin{equation}
    \label{eq:d} 
    \left(\begin{matrix}
    0 & -1 & 0\\
    -1 & 4 & -1\\
    0 & -1 & 0
    \end{matrix}\right).
\end{equation}
We construct an $N\times N$ block-Toeplitz matrix $\mathbf{D}_\ell$ with $N$ Toeplitz blocks; see~\cite{henrot2012fast} for details. Imposing periodic boundary conditions on $\mathbf{D}_\ell$, it can be reformulated as a block circulant matrix with circulant blocks, a structure denoted as \emph{circulant-block-circulant} (CBC). This property allows to diagonalize $\mathbf{D}_\ell$ with 2D Fourier transforms. 
This leads to matrix $\mathbf{D}$ in~\eqref{eq:object} with block-diagonal structure:
\begin{equation}
    \label{eq:D}
    \mathbf{D}  = \begin{pmatrix}
    \mathbf{D}_{1} & \bf 0 &\ldots &\bf 0 \\
    \bf 0  &\ddots &\ddots &\vdots\\
    \vdots &\ddots &\ddots &\bf 0\\
    \bf 0 & \ldots & \bf 0 & \mathbf{D}_{B}
\end{pmatrix}.
\end{equation}

Matrix $\mathbf{E}$ cannot be block-diagonal since it operates across channels. A typical choice is to penalize the variation between two adjacent channels with a first-order derivative filter \cred{$\mathcal{E}_0 = [1 -1]$} along the spectral dimension. The convolution matrix, of size $(B-1)\times B$, is then given by:
\begin{equation}
    \label{eq:E_0}
    \mathbf{E}_0  = \begin{pmatrix}
    -1 & 1 & 0 & \ldots & 0 \\
    0  & -1 & 1 & \ddots &\vdots\\
    \vdots &\ddots &\ddots &\ddots &0\\
    0 & \ldots & 0 & -1 & 1
    \end{pmatrix}.
\end{equation}
This yields:
\begin{equation}
    \label{eq:E}
    \mathbf{E}  = \mathbf{E}_0 \otimes \mathbf{I}_N
\end{equation}
where $\otimes$ denotes the Kronecker product of two matrices and $\mathbf{I}_N$ is the $N\times N$ identity matrix. 
Although $\mathbf{D}$ and $\mathbf{E}$ are large matrices of size ${BN\times BN}$, they are not explicitly stored or used in practice. As can be seen in Subsection~\ref{sec:v},  $\|\mathbf{D}(\mathbf{x} - \tilde{\mathbf{x}})\|^{2}$ and $\|\mathbf{E}(\mathbf{x} - \tilde{\mathbf{x}})\|^{2}$ are actually computed in an efficient manner with Fourier transforms. 

\section{Numerical Optimization}
\label{sec:optimization}

We shall now introduce the algorithm to estimate $\mathbf{X}$ by minimizing the objective function~\eqref{eq:object}. First, we use
a variable splitting technique, i.e., the Half Quadratic Splitting (HQS)~\cite{geman1995nonlinear} algorithm, to decompose the optimization problem into iterative sub-problems. One sub-problem related to data fidelity terms is solved based on a fast Sylvester analytical solver. Another sub-problem involving regularizers is solved efficiently using 2D discrete Fourier transform.  

\vspace{-3mm}
\subsection{Variable splitting based on HQS} 
\vspace{-1mm}
HQS is employed to decouple the data fidelity terms and the regularization terms in~\eqref{eq:object}. By introducing an auxiliary variable $\mathbf{V}\in\mathbb{R}^{B\times N}$, minimization of~\eqref{eq:object} can be reformulated as:
\begin{equation}
\label{eq:objective2}
\begin{split}
&\hat{\bf X} = \mathop{\min}_{{\bf X}}  \, \|\mathbf{Y} - \mathbf{XBS}\|_{F}^{2} +
\|\mathbf{Z} - \mathbf{RX}\|_{F}^{2}  + \mu\|\mathbf{D}(\mathbf{v} - \tilde{\mathbf{x}})\|^{2} \\ &\qquad \qquad +\nu\|\mathbf{E}(\mathbf{v} - \tilde{\mathbf{x}})\|^{2} \qquad \quad \  {\rm s.t.} \qquad \quad {\bf V} = {\bf X} .
\end{split}
\end{equation}
where $\mathbf{v}\in\mathbb{R}^{BN}$ denote the vector obtained by stacking the columns of the matrice $\bf{V}$.
The augmented Lagrangian function is given by:
\begin{equation}
\label{eq:Lag}
\begin{aligned}
\mathcal{L}_{\rho}({\bf X, V}) = \,  \,  & \|\mathbf{Y} - \mathbf{XBS}\|_{F}^{2} +
\|\mathbf{Z} - \mathbf{RX}\|_{F}^{2} +{\rho}\|{\bf X-V}\|_{F}^{2} \\ & + \nu\|\mathbf{E}(\mathbf{v} - \tilde{\mathbf{x}})\|^{2} + \mu\|\mathbf{D}(\mathbf{v} - \tilde{\mathbf{x}})\|^{2}
\end{aligned}
\end{equation}
where $\rho$ is a positive penalty parameter. HQS method then minimizes~\eqref{eq:Lag} via the following steps:
\begin{equation}
	\label{eq:stepa} {\bf X}_{k+1} = \mathop{\min}_{{\bf X}}\|\mathbf{Y} - \mathbf{XBS}\|_{F}^{2} + \|\mathbf{Z} - \mathbf{RX}\|_{F}^{2} + {\rho}\|{\bf X-V}_k\|_{F}^{2}
\end{equation}
\begin{equation}
	\label{eq:stepb} {\bf v}_{k+1} = \mathop{\min}_{{\bf v}} {\rho}\|{\bf x}_{k+1}-\mathbf{v}\|^{2} + \mu\|\mathbf{D}(\mathbf{v} - \tilde{\mathbf{x}})\|^{2} + \nu\|\mathbf{E}(\mathbf{v} - \tilde{\mathbf{x}})\|^{2}
\end{equation}
We can observe that the fidelity terms and the regularization terms are decoupled in the two sub-problems~\eqref{eq:stepa} and~\eqref{eq:stepb}. Now we can address the above
minimization problems in an efficient manner by iteratively minimizing with respect
to $\mathbf{X}$ and $\mathbf{v}$, independently. These two steps are as follows.

 \vspace{-3mm}
\subsection{Optimization \cred{w.r.t.} $\mathbf{X}$} 
 \vspace{-1mm}
\renewcommand{\algorithmicrequire}{ \textbf{Input:}} 
\renewcommand{\algorithmicensure}{ \textbf{Output:}} 
\begin{algorithm}[!t]
	\caption{Solving the Sylvester equation w.r.t. ${\bf X}_{k+1}$}
	\label{alg_1}
	\begin{algorithmic}
		\REQUIRE $\mathbf{Y}$, $\mathbf{Z}$, $\mathbf{B}$, $\mathbf{S}$, $\mathbf{R}$,  ${\mathbf{V}}_k$, $\rho$.\\
		\ENSURE ${\mathbf{X}}_{k+1}$.\\
		\STATE Initialize $\mathbf{C_1}, \mathbf{C_2}, \mathbf{C_3}$ via~\eqref{eq:C}
		\STATE Compute the eigen-decomposition of $\mathbf{B}$ as ${\mathbf{FDF}^{H}}$
		\STATE $\overline{\mathbf{D}} = \mathbf{D}(\mathbf{1}_s\otimes \mathbf{I}_n)$
		\STATE Compute the eigen-decomposition of $\mathbf{C_1}$ as $\mathbf{C_1} = {\mathbf{Q\Lambda Q}^{-1}}$
		\STATE $\overline{\mathbf{C_3}} = \mathbf{Q}^{-1}\mathbf{C_3}\mathbf{F}$
		\STATE Compute auxiliary matrix $\overline{\mathbf{X}}$ channel by channel
		\FOR{$\ell=1$ to $B$}
		\STATE $\overline{\mathbf{X}}_{\ell}=\lambda_{\ell}^{-1}({\overline{\mathbf{C_3}}})_i - \lambda_{\ell}^{-1}({\overline{\mathbf{C_3}}})_k\overline{\mathbf{D}}(\lambda_{\ell}s\mathbf{I}_n + \sum\limits_{t=1}^{s} \mathbf{D}_t^2)\overline{\mathbf{D}}^{H}$
		\ENDFOR
		\STATE ${\mathbf{X}}_{k+1} = \mathbf{Q}\overline{\mathbf{X}}\mathbf{F}^{H}$		
	\end{algorithmic}
\end{algorithm}
Let us assume that the blurring matrix $\mathbf{B}$ has an CBC structure. Under this widely accepted assumption, matrix $\mathbf{B}$ can be decomposed as follows: $\mathbf{B} = \mathbf{FDF}^{H}$ with $\mathbf{F}\in\mathbb{R}^{N\times N}$ the DFT matrix, $\mathbf{D}\in\mathbb{R}^{N\times N}$ a diagonal matrix, and $^H$ the conjugate transpose.

To solve sub-problem~\eqref{eq:stepa}, we set the gradient of the objective function~\eqref{eq:stepa} \cred{w.r.t.} $\mathbf{X}$ to zero. Thus, ${\bf X}_{k+1}$ is the solution of the Sylvester equation:
\begin{equation}\label{eq:Sylvester}
\mathbf{C_1}{\mathbf{X}}_{k+1} + {\mathbf{X}}_{k+1}\mathbf{C_2} = \mathbf{C_3}
\end{equation}
where 
\begin{equation}
\label{eq:C}
\begin{split}
\mathbf{C_1} &= \mathbf{R}^{T}\mathbf{R}+\mu \mathbf{I}_{B}\\
\mathbf{C_2} &= (\mathbf{BS})(\mathbf{BS})^{T}\\
\mathbf{C_3} &= \mathbf{R}^{T}\mathbf{Z} + \mathbf{Y}(\mathbf{BS})^{T} + \rho{\mathbf{V}_k}
\end{split}
\end{equation}
and $\mathbf{I}_{B}$ is the identity matrix of size $B\times B$. 

According to the conclusion in~\cite{bartels1972solution}, the Sylvester equation in~\eqref{eq:Sylvester} has a unique solution when an arbitrary sum of the eigenvalues of $\mathbf{C_1}$ and $\mathbf{C_2}$ is not equal to zero. Matrix $\mathbf{C_1}$ is positive-definite since $\mathbf{R}^{T}\mathbf{R}$ and $\mathbf{I}_{B}$ are both positive-definite matrices, and $\mathbf{C_2}$ is positive semi-definite. Thus, any sum of the eigenvalues of $\mathbf{C_1}$ and $\mathbf{C_2}$ is larger than zero, ensuring the uniqueness of the solution of~\eqref{eq:Sylvester}. For a fast algorithm solving~\eqref{eq:Sylvester}, the interested reader can refer to~\cite{wei2015fast}. The steps are summarized in Algorithm~\ref{alg_1}.
\renewcommand{\algorithmicrequire}{ \textbf{Input:}} 
\renewcommand{\algorithmicensure}{ \textbf{Output:}} 
\begin{algorithm}[!t]
	\caption{HSI super-resolution accounting for spectral-spatial gradient deviation}
	\label{alg_2}
	\begin{algorithmic}
		\REQUIRE $\mathbf{Y}$, $\mathbf{Z}$, $\mathbf{B}$, $\mathbf{S}$, $\mathbf{R}$,  $\tilde{\mathbf{X}}$, $\mu$, $\nu$, $\rho$\\
		\ENSURE $\hat{\mathbf{X}}$\\
		\STATE Initialize ${\bf V}_0 = \tilde{\mathbf{X}}$, $k=0$, $\mu' = \mu / \rho$, $\nu' = \nu / \rho$
		\WHILE{stopping criterion is not met and $k < K$}
		\STATE Update ${\bf X}_{k+1}$ via Algorithm~\ref{alg_1}
		\STATE For $\ell = 1,\ldots, B$, compute 2D DFTs of data ${\mathcal{X}}_{k+1, \ell}$ and $\tilde{\mathcal{X}}_{\ell}$
		\STATE For each $\bf f$, compute $\underline{\mathcal{V}}_{\,\ell}(\mathbf{f})$ using~\eqref{eq:solution} and~\eqref{eq:tf}
		\STATE For $\ell = 1,\ldots, B$, compute ${\mathbf{v}}_{k+1, \ell}$ from $\underline{\mathcal{V}}_{\,\ell}$ via inverse 2D DFTs
		\STATE $k = k + 1$
		\ENDWHILE	
		\STATE $\hat{\mathbf{X}} = \mathbf{X}_K$
	\end{algorithmic}
\end{algorithm}
\renewcommand{\baselinestretch}{1.1}
\begin{table*}[]
	\centering
	\caption{Averaged RMSE, PSNR, SAM, ERGAS and SSIM of different methods on the CAVE and Harvard data sets.}
	\begin{tabular}{c|ccccc|ccccc}
		\hline \hline
		\multirow{2}{*}{Methods} & \multicolumn{5}{c|}{CAVE data set}              
		& \multicolumn{5}{c}{Harvard data set}         \\ \cline{2-11}
		& RMSE & PSNR & ERGAS & SAM & SSIM & RMSE & PSNR & ERGAS & SAM & SSIM \\ \hline \hline
		
		UAL & 1.854 & 44.656 & 0.196 & 4.33 & 0.9910 & 1.833 & 45.807 & 0.323 & 3.58 & 0.9832 \\
		UAL + Ours & \textbf{1.587} & \textbf{45.939} & \textbf{0.171} & \textbf{4.08} & \textbf{0.9917} & \textbf{1.784}	& \textbf{46.034} &\textbf{ 0.316}	& \textbf{3.54}	& \textbf{0.9833}\\
		\hline
		NSSR & 2.236 & 43.439 & 0.244 & 5.22 & 0.9849 & 1.874 & 45.540 & 0.363 & 3.73 & 0.9821\\
		NSSR + Ours & \bf 2.068 & \bf 44.044 & \bf 0.230 &  \textbf{5.19} & \bf0.9854 & \textbf{1.844} & \textbf{45.649} &  \textbf{0.357} & \textbf{3.69} & \textbf{0.9822}\\
		\hline
		LTTR & 2.300 & 43.277 & 0.249 & 5.50 & 0.9848 & 1.914 & 45.251 & 0.375  & 3.81 & 0.9813 \\
		LTTR + Ours & \textbf{2.235} & \textbf{43.613 }& \textbf{0.243} & \textbf{5.27} & \textbf{0.9851} & \textbf{1.887} & \textbf{45.392} &  \textbf{0.374} & \textbf{3.77} & \textbf{0.9915} \\
		\hline \hline
	\end{tabular}
	\label{tab_1}
	\vspace{-2mm}
\end{table*}

 \vspace{-3mm}
\subsection{Optimization \cred{w.r.t.} $\mathbf{v}$} 
 \vspace{-1mm}
\label{sec:v}
Following~\cite{henrot2012fast} and recalling the notation system $(\mathcal{X}, \mathbf{X}, \mathbf{x})$ for other images as described at the end of Section~\ref{sec:introduction}, we can rewrite the objective function in~\eqref{eq:stepb} in 3D image domain with a sum running over spectral channels:
\begin{equation}\label{eq:channel}
\begin{aligned}
\min_{\mathcal{V}}\sum_{\ell = 1}^{B} \Big( \rho\|{\mathcal{X}}_{k+1, \ell}-\mathcal{V}_{\ell}\|_{F}^{2} &+ \mu\|\mathcal{D}_{\ell}{*}_{2D}(\mathcal{V}_{\ell} - \tilde{\mathcal{X}}_{\ell})\|_{F}^{2} \\
& + \nu\|[\mathcal{E}_0{*}_{1D}(\mathcal{V} - \tilde{\mathcal{X}})]_{\ell}\|_{F}^{2}\Big)
\end{aligned}
\end{equation}
The operator ${*}_{2D}$ denotes 2D convolution in the spatial domain while ${*}_{1D}$ represents 1D convolution across spectral channels. Using \cred{$L\times W$} 2D DFT \cred{in the spatial domain, and denoting the Fourier transformed quantities with underlined symbols of the spatial frequency variable $\bf f$,} Plancherel theorem allows to rewrite~\eqref{eq:channel} as:
\begin{equation}\label{eq:fourier}
\begin{aligned}
\min_{\underline{\mathcal{V}}}\sum_{\ell = 1}^{B}\Big(\rho\|\underline{\mathcal{X}}_{\, k+1, \ell}-\underline{\mathcal{V}}_{\,\ell}\|_{F}^{2} &+ \mu\|\,\underline{\mathcal{D}}_{\,\ell}\odot (\underline{\mathcal{V}}_{\,\ell} - \underline{\tilde{\mathcal{X}}}_{\,\ell})\|_{F}^{2} \\
& + \nu\| [\mathcal{E}_0{*}_{1D}(\underline{\mathcal{V}} - \tilde{\underline{\mathcal{X}}})]_{\ell}\|_{F}^{2}\Big)
\end{aligned}
\end{equation}
with $\odot$ the Hadamard product. Given a channel $\ell$, $[\mathcal{E}_0{*}_{1D}(\mathcal{V} - \tilde{\mathcal{X}})]_{\ell}$ is a sum of images $( \mathcal{V} - \tilde{\mathcal{X}})_{\ell'}$ weighted by the coefficients in $\mathcal{E}_0$. As the DFT is not calculated across channels, $[\mathcal{E}_0{*}_{1D}(\underline{\mathcal{V}} - \tilde{\underline{\mathcal{X}}})]_{\ell}$ is also a sum of images $(\underline{\mathcal{V}} - \tilde{\underline{\mathcal{X}}})_{\ell'}$ with the same coefficients. 

Since the convolution in $\mathcal{E}_0{*}_{1D}(\underline{\mathcal{V}} - \tilde{\underline{\mathcal{X}}})$ is performed in the spectral channel domain, the minimization problem~\eqref{eq:fourier} can be separated into independent minimisation sub-problems  \cred{w.r.t.} each spatial frequency $\bf f$. The optimization problem in~\eqref{eq:stepb} can be decomposed into a set of independent least square problems, each corresponding to a point $\bf f$ in the spatial frequency domain:
\begin{equation}
\label{eq:f}
\mathop{\min}_{{\bf v}_{\mathbf{f}}} \|{\bf x}_{k+1, \mathbf{f}}-\mathbf{v}_{\mathbf{f}}\|^{2} + \mu'\|\Delta_\mathcal{D}(\mathbf{f})(\mathbf{v}_{\mathbf{f}} - \tilde{\mathbf{x}}_{\mathbf{f}})\|^{2} + \nu'\|\mathbf{E}_0(\mathbf{v}_{\mathbf{f}} - \tilde{\mathbf{x}}_{\mathbf{f}})\|^{2}
\end{equation}
with $\mu' = \mu / \rho$ and $\nu' = \nu / \rho$. The complex vectors ${\bf v}_{\mathbf{f}}, \tilde{\mathbf{x}}_{\mathbf{f}}, {\bf x}_{k+1, \mathbf{f}}$ and the complex diagonal matrix $\Delta_\mathcal{D}$ are defined as follows:
\begin{equation}
\label{eq:complex}
\begin{split}
{\bf v}_{\mathbf{f}} &= \left\{\underline{\mathcal{V}}_{\,\ell}(\mathbf{f}), \ell = 1, \ldots, B \right\}\\
\tilde{\mathbf{x}}_{\mathbf{f}} &= \{\underline{\tilde{\mathcal{X}}}_{\,\ell}(\mathbf{f}), \ell = 1, \ldots, B \}\\
{\bf x}_{k+1, \mathbf{f}} &= \{ \underline{\mathcal{X}}_{\, k+1, \ell}(\mathbf{f}), \ell = 1, \ldots, B \}\\
\Delta_\mathcal{D}(\mathbf{f}) &= \mathrm{diag}\left\{\underline{\mathcal{D}}_{\,\ell}(\mathbf{f}), \ell = 1, \ldots, B\right\}\\
\end{split}
\end{equation}
For each $\bf f$, the solution of~\eqref{eq:f} can be computed as:
\begin{equation}
\label{eq:solution}
{\bf v}_{\mathbf{f}} = \mathbf{T_f}^{-1}({\bf x}_{k+1, \mathbf{f}}+\mu'\Delta_\mathcal{D}(\mathbf{f})^*\Delta_\mathcal{D}(\mathbf{f})\tilde{\mathbf{x}}_{\mathbf{f}} + \nu' \mathbf{E}_0^*\mathbf{E}_0\tilde{\mathbf{x}}_{\mathbf{f}})
\end{equation}
where $^*$ denotes the complex conjugate and $\mathbf{T_f}$ is the real tri-diagonal matrix of size $B\times B$ given by:
\begin{equation}
\label{eq:tf}
\mathbf{T_f}=(\mathbf{I}_B + \mu' \Delta_\mathcal{D}(\mathbf{f})^*\Delta_\mathcal{D}(\mathbf{f}) + \nu' \mathbf{E}_0^*\mathbf{E}_0)
\end{equation}
Finally, we can obtain ${\bf v}_{k+1}$ by separately calculating the inverse 2D DFT of each $\underline{\mathcal{V}}_{\,\ell}$ with $\ell = 1, \ldots, B$.
This procedure is summarized in Algorithm~\ref{alg_2}.

\section{Experiments}
\label{sec:exp}

We shall now validate the proposed strategy with experimental results. The code is made available at github.com/xiuheng-wang.

Two public HSI data sets, namely, the CAVE data set~\cite{yasuma2010generalized} and the Harvard data set~\cite{chakrabarti2011statistics}, were used for our experiments. The CAVE data set is composed of 32 HSIs with a spatial dimension $512\times 512$, and 31 channels in the spectral domain, covering the visible spectrum from 400 nm to 700 nm. The Harvard data set contains 50 HSIs consisting of $1392\times 1040$ pixels in the spatial domain, and 31 channels ranging from 420 nm to 720 nm. For the Harvard data set, the top left $1024\times 1024$ pixels were cropped and extracted. 

The HSIs of the two data sets were scaled to range $[0, 1]$ and served as the ground truth for ${\bf X}$. The LR HSI ${\bf Y}$ was generated according to~\eqref{eq:degradation} where $\mathbf{B}$ is a uniform blurring operator over non-overlapping blocks of size $32\times 32$, and $\mathbf{S}$ is a down-sampling operator with the down-sampling factor $s = 32$. The HR RGB $\mathbf{Z}$ image was obtained with~\eqref{eq:degradation}, with $\mathbf{R}$ the response of a Nikon D700 camera.

\begin{figure}[tp]
	\centering
	\includegraphics[scale=0.36]{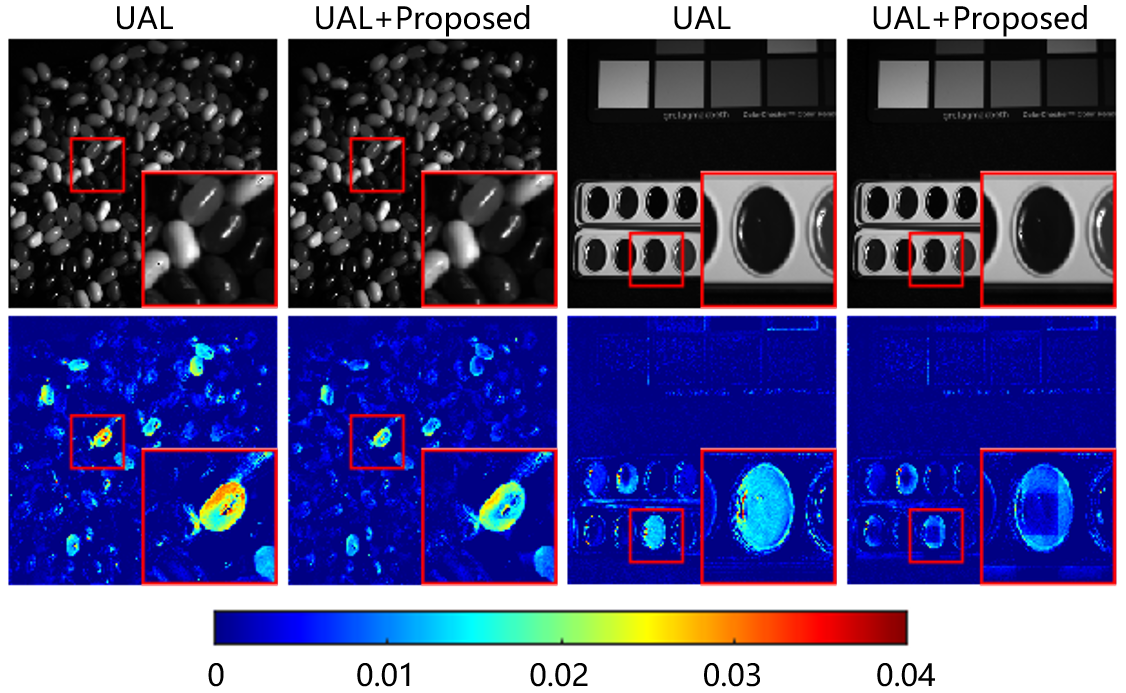}
	\caption{Reconstructed images and corresponding error maps of two images from the CAVE data set in the 540 nm band. }
	\label{fig:comparison}
	\vspace{-3mm}
\end{figure}

We used a state-of-the-art deep learning method UAL described in~\cite{zhang2020unsupervised} to calculate the deep prior $\tilde{\bf X}$ for each ${\bf X}$. We set the hyper-parameters as follows: $\mu = 0.05$, $\nu = 0.001$ in~\eqref{eq:object}, and $\rho = 0.001$ in~\eqref{eq:Lag}. The number of iterations $K$ in Algorithm~\ref{alg_2} was set to $20$, which was sufficient to ensure convergence. To assess the quality of reconstructed images, we considered the following metrics: the root mean-square error (RMSE), the peak-signal-to-noise-ratio (PSNR), the spectral angle mapper (SAM)~\cite{yuhas1992discrimination}, the error of relative global adimensional synthesis (ERGAS)~\cite{wald2000quality} and the structural similarity (SSIM)~\cite{wang2004image}. Table~\ref{tab_1} reports the performance of the UAL super-resolution algorithm~\cite{zhang2020unsupervised}, and of our algorithm which combines the UAL and the degradation model inversion (Algorithm~\ref{alg_2}). It can be observed that our algorithm significantly improved the performance of the UAL. 
Figure~\ref{fig:comparison} confirms this observation by showing that our approach produced smaller reconstruction errors than the UAL. 

For comparison purpose, we also considered other state-of-the-art super-resolution algorithms than the UAL to produce priors $\tilde{\bf X}$: the NSSR~\cite{dong2016hyperspectral} based on sparse decompositions, and the LTTR~\cite{dian2019learning} based on tensor factorizations. The rational was to show that coupling our approach with these algorithms improves their performance as it makes use of the linear degradation model and exploits the spectral-spatial smoothness of HSIs. Algorithm~\ref{alg_2}
was setup as described above. To setup NSSR and LTTR, we used the codes provided by their authors and fine-tuned all
parameters to achieve the best super-resolution performance. 
Table~\ref{tab_1} confirms that our approach allowed us to improve the performance of both NSSR and LTTR. The best performance was however achieved by using, with our algorithm, the deep priors provided by the UAL.

\section{Conclusion}
\label{sec:con}
In this paper, we introduced an HSI super-resolution method which makes use of a degradation model in the data-fidelity  term of the objective function and, on the other hand, utilizes the spectral-spatial gradient deviation of latent HSIs and the output of a convolutional neural network as a deep prior regularizer. Experiments showed the performance improvement achieved with this strategy compared with state-of-the-art methods. 

\clearpage
\newpage
\balance

\label{sec:refs}

\bibliographystyle{IEEEbib}
\bibliography{strings}

\end{document}